\documentclass{article}
\usepackage{pst-all}
\usepackage{pstcol}
\usepackage{amsmath}
\usepackage{amssymb}
\usepackage[dvips]{graphicx}
\usepackage{graphicx}
\begin{document}
\title{\textbf{Quantum Gravity Effects in Rotating Black Holes}\footnote{Talk given by E.T. at the Eleventh Marcel Grossmann 
Meeting on General Relativity, Freie Universit\"at Berlin: July 23 - 29, 2006.}}
\author{M. Reuter, E. Tuiran \\Institute of Physics, University of Mainz \\ D-55099 Mainz, Germany}
\maketitle
\section{Introduction}\label{intro}
The effective average action has been used for
detailed studies of the nonperturbative renormalization behavior of Quantum Einstein Gravity, in particular in 
the context of the asymptotic safety scenario \cite{mr,as}. As a first application of the 
scale dependent Newton constant
derived in \cite{mr}, quantum corrections to the Schwarzschild spacetime were discussed in \cite{Bonanno-Reuter}. 
Indications were found that, due to quantum effects, the Hawking evaporation 
process stops once the mass of the black hole is of the order 
of the Planck mass \cite{Bonanno-Reuter,evap}. In this note we report on some aspects of the
corresponding analysis for Kerr black holes \cite{long}.
\section{Renormalization Group Improvement}
The (truncated) renormalization group (RG) equation for the average action provides us with a running Newton constant
$G\left(k\right)$ where the mass parameter $k$ sets the scale of the ``coarse graining'' which has been performed. 
Technically it is 
implemented as an infrared cutoff in the underlying functional integral over all metrics. In the improvement approach
of \cite{Bonanno-Reuter} one tries to relate $k$ to the geometrical properties of the system under consideration. 
More concretely, one sets up a correspondence $k=k\left(P\right)$ between scales and spacetime points $P$. 
A still rather general
ansatz for this correspondence is $k \propto 1/d\left(P\right)$, where 
$d\left(P\right)=\int_{\cal{C}}\sqrt{\left| ds^{2}\right|}$ is the proper length of a spacetime curve $\cal{C}$
related to $P$ which is computed 
with respect to the classical metric. This ansatz is manifestly diffeomorphism invariant, and thanks to
its nonlocal character it is potentially capable of mimicking certain (not explicitly known) nonlocal terms in the average
action. In \cite{long} various choices for $\cal{C}$ are discussed, for instance a radial path from the center of the
black hole to $P$. Using standard Boyer-Lindquist (BL) coordinates, $d\left(P\right)$ becomes a function 
$d\left(r,\theta\right)$. It turns out \cite{long} that within the expected domain of reliability of this approach
 the $\theta-$dependence of the invariant distance is inessential and $d \approx d\left(r\right)$ depends on the 
radial coordinate only. Asymptotically, $d\left(r\to\infty\right) \approx r$. \newline
For a ``semi-quantitative'' analysis we used the approximation for $G\left(r\right)$ 
given by $G\left( k\right)=G_{0} / \left(1 + w G_{0} k^2 \right)$ \cite{Bonanno-Reuter}. 
It entails the position dependent Newton constant 
$G\left( r\right)=G_{0}d^2\left( r\right)/ \left(d^2\left( r\right)+\bar{w}G_{0}\right)$. 
Here $w$ and $\bar{w}$ are positive constants, 
and $G_0 \equiv M^{-2}_{\rm Pl}$ is the standard Newton constant. The
``RG improvement'' consists in substituting $G_0 \rightarrow G\left(r\right)$ in the classical Kerr solution. The 
resulting metric of the RG improved Kerr spacetime reads, in BL coordinates,
\begin{eqnarray*}
ds_{\rm imp}^2=-\left(\Delta\rho^{-2}\right)\left[dt-a \sin^2\theta d\varphi\right]^2+
\left(\rho^{-2}\sin^2 \theta\right) \left[\left(r^2+a^2\right)d\varphi - a dt\right]^2
+\\ \left(\rho^2\Delta^{-1}\right) dr^2+\rho^2 d\theta^2.
\end{eqnarray*}
The general structure of this metric, as well as the abbreviations $\rho^2 \equiv r^2+a^2 \cos^2 \theta$, 
$a \equiv J/M$, are as in the classical case. The only place where $G\left(r\right)$ appears is in 
$\Delta \equiv r^2-2G\left(r\right)M r +a^2$.
\section{Critical Surfaces of the Improved Kerr Metric}
The spacetime described by $ds_{\rm imp}^2$ has two infinite redshift surfaces $\left(g_{00}=0\right)$ at 
radii $r=r_{S_{\pm}}\left(\theta\right)$ given by $r^2-2G\left(r\right) M r + a^2\cos^2\theta =0$. 
We denote them by $S_{\pm}$. The outer one, $S_+$, is the static limit surface. 
Furthermore, the spacetime has two event horizons ($g^{rr}=0,\ \Delta=0$) 
with radii $r=r_{\pm}$ to be obtained from $r^2-2G\left(r\right) M r + a^2 =0$. We denote the inner (outer) horizon by 
$H_{-}$ ($H_{+}$). In Fig. 1 we plot the radii $r_{\pm}$ and $r_{S_{\pm}}$ ($\theta=\frac{\pi}{2}$) in the equatorial 
plane for the approximation $d\left(r\right)=r$, and we compare them to the classical values ($\bar{w}$=0). The
upper and lower branches of the curves correspond to $S_{+}$, $H_{+}$ and $S_{-}$, $H_{-}$, respectively. We observe that 
for small enough $M$ the horizons coalesce and then disappear, and similarly for $S_{\pm}$ at even lower masses. The 
coalescence of $H_{\pm}$ and $S_{\pm}$ occur for $M$ of the order of $M_{\rm Pl}$ where the applicability of the method 
becomes questionable. It can be safely applied for $r \gg l_{\rm Pl}$ if $M \gg m_{\rm Pl}$ and $a \ll MG_0$ 
where the quantum corrections are small.
\centerline{}
\begin{center}
\begin{pspicture}(-1,6)(4,1)
\includegraphics{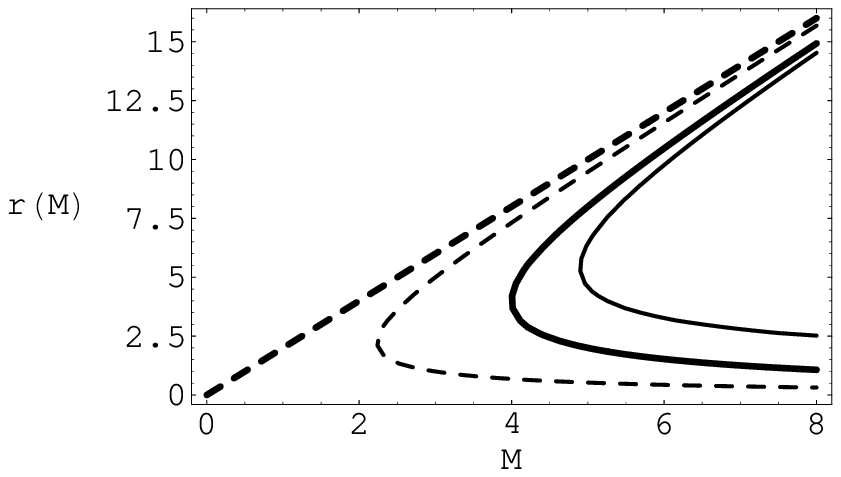}
\rput[l](-5,6){$a=4$}
\rput[l](-3.7,1.5){\small{Fig. 1.}}
\rput[l](-9.45,1.1){\small{Radial coordinates of the critical surfaces at the equatorial plane vs. mass in Planck units}}
\rput[l](-9.45,.7){\small{and their improved counterparts. Dashed lines represent the static limit surfaces 
$S_{\pm}$, solid}}
\rput[l](-9.45,0.3){\small{lines the event horizons $H_{\pm}$. The thicker lines correspond to the classical surfaces.}}
\end{pspicture}
\end{center}
\centerline{}
\section{Antiscreening and Smarr's Formula}
Since the improved Kerr metric is known explicitly, we can compute its Einstein tensor and write it in the form 
$G_{\mu\nu}=8\pi G_0 T^{\rm{eff}}_{\mu\nu}$, thus defining an effective energy-momentum tensor 
$T^{\rm{eff}}_{\mu\nu}$ for the quantum fluctuations which drive the renormalization group evolution. Nevertheless,
improved vacuum black holes are in many respects quite different from classical ones in presence of matter. The reason 
is that $T^{\rm{eff}}_{\mu\nu}$ does not have any of the standard positivity properties which are crucial in black hole 
thermodynamics, for instance. 
Corrections to the mass $M_{\rm H}$ and angular momentum $J_{\rm H}$ of the Kerr black hole coming from the 
``pseudo-matter'' described by $T^{\rm{eff}}_{\mu\nu}$ can be calculated by performing 
the Komar integrals at the event horizon:
$M_{\rm H}=-\left(8\pi G_0\right)^{-1}\oint \nabla^{\alpha}t^{\beta}dS_{\alpha\beta}$\ ,\ 
$J_{\rm H}=\left(16\pi G_0\right)^{-1}\oint \nabla^{\alpha}\phi^{\beta}dS_{\alpha\beta}$. 
Here $t^{\beta}$ and $\phi^{\beta}$ are the Killing vectors associated to stationarity and axial symmetry, respectively. 
One finds:
\begin{eqnarray}
M_{\rm H} &=&\frac{MG\left(r_+\right)}{G_0}
\left[1-\arctan\left(\frac{a}{r_+}\right)\frac{G'(r_+)\left(r^2_+ + a^2\right)}{aG(r_+)}\right]  \label{M_H Res}
\end{eqnarray}
\begin{eqnarray}
J_{\rm H}&=&\frac{JG(r_+)}{G_0}
+\frac{M^2r^2_+G'(r_+)G(r_+)}{G_0a}\left[1-\frac{2MG(r_+)}{a}\arctan\left(\frac{a}{r_+}\right)\right] 
\end{eqnarray}
For the case of the mass, (\ref{M_H Res}) tells us that, due to quantum fluctuations, 
the classical mass $M$ is decreased to a value $M_{\rm H}<M$ for every possible 
running of the Newton's constant \cite{long}. This can be interpreted due to the \textit{antiscreening} character 
of quantum gravity \cite{mr}. Remarkably enough, Smarr's formula still holds in its classical form
$M_{\rm H} = 2\Omega_{\rm H} J_{\rm H}+ \kappa A /\left(4\pi G_0\right)$.
For the quantum corrected black hole, the horizon's angular frequency, surface gravity, and area are given by 
$\Omega_{\rm H} = a/\left(r^2_++a^2\right)$, 
$\kappa=\left(r_+-2M\right)\left[G(r_+)+r_+G'(r_+)\right]/\left(r^2_++a^2\right)$ and
$A=4\pi\left(r^2_++a^2\right)$.
The classical appearance of these formulas (except for the $G'$-term in $\kappa$) is deceptive: The radius 
$r_+ \equiv r_+\left(a,M\right)$ depends on the parameters of the black hole, and this relationship is modified 
by the renormalization effects.
\section{Summary}
We explained how quantum gravity effects in the spacetime of rotating black holes can be taken into account by a RG 
improvement of the classical Kerr solution. We discussed the Black hole's critical surfaces as well as the 
``gravitational dressing'' of its mass and angular momentum. 
Further properties of the improved Kerr black hole, 
in particular its thermodynamics and Penrose process, will be described in ref. \cite{long}.

\end{document}